\documentclass{appolb}
\usepackage{epsfig}

\begin{document}
\title{Are the contemporary financial fluctuations \\
sooner converging to normal?}

\author{S. Dro\.zd\.z$^{1-4}$, J. Kwapie\'n$^{1,2,5}$, F.
Gr\"ummer$^{1}$, F. Ruf$^{6}$, and J. Speth$^{1,5}$
\address{$^1$ Institut f\"ur Kernphysik, Forschungszentrum J\"ulich,
D-52425 J\"ulich, Germany \\
$^2$ Institute of Nuclear Physics, PL--31-342 Krak\'ow, Poland \\
$^3$ Institute of Physics, University of Rzesz\'ow, PL--35-310 Rzesz\'ow, 
Poland \\
$^4$ Physikalisches Institut, Universit\"at Bonn, D--53115 Bonn, Germany 
\\
$^5$ Special Research Center for the Subatomic Structure of Matter, 
University of Adelaide, SA 5005, Australia \\
$^6$ West LB International S.A., 32-34 bd Grande-Duchesse Charlotte,
L-2014 Luxembourg}
}
\maketitle
\begin{abstract}
Based on the tick-by-tick price changes of the companies from the U.S.
and from the German stock markets over the period 1998-99 we reanalyse
several characteristics established by the Boston Group for the U.S.
market in the period 1994-95, which serves to verify their space and
time-translational invariance. By increasing the time scales, in the
region covered by the data, we find a significantly more accelerated
departure from the power-law $(\alpha \approx 3)$ asymptotic behaviour of
the distribution of returns towards a Gaussian, both for the U.S. as well
as for the German stock markets. In the latter case the crossover is even
faster. Consistently, the corresponding autocorrelation functions of
returns and of the time averaged volatility also indicate a faster loss
of memory with increasing time. This route towards efficiency, as seen in
a fixed time scale, may reflect a systematic increase of the quality of
information processing when going from past to present.
\end{abstract}
  
\section{Introduction}

Besides its obvious practical implications studying the nature of
financial fluctuations proves extremely inspiring and productive for
fundamental reasons~\cite{burda}. The related contributions by
Bachelier~\cite{Bachelier} and by Mandelbrot~\cite{Mandelbrot}, and broad
scientific consequences of these contributions, provide immediate
examples. In financial dynamics, even though somewhat opposite, the two
corresponding scenarios of uncorrelated random Gaussian~\cite{Bachelier},
versus L\'evy stable~\cite{Mandelbrot} fluctuations, turn out to be
taking part and leaving their imprints. As documented by Stanley and
collaborators~\cite{Mantegna,Gopi,Plerou}, the central part of the
distribution of returns falls within the L\'evy stable regime, while
larger fluctuations are governed by a power law with an exponent $\alpha
\approx 3$, well outside the L\'evy stable regime. At the same time the
autocorrelation function for returns sampled at short time scales drops
down very quickly and after about 20 min it reaches the noise level.
Consequently, because of the central limit theorem, the convergence to a
Gaussian distribution on longer time scales is expected. Quite
surprisingly, such a convergence has been
shown~\cite{Mantegna,Gopi,Plerou} to be extremely slow. In fact, for
returns of up to approximately 4 days, the functional form of their
distribution even is retained for both, the individual
companies~\cite{Plerou} as well as for the global stock market
index~\cite{Gopi}. Based on this analysis a visible crossover to a
Gaussian takes place only after about 16 days. The volatility
autocorrelation function, on the other hand, decays very slowly with
time, largely according to a power law, and remains positive for many
months. These higher order correlations can thus be considered
responsible for such an ultraslow convergence to a Gaussian. These, at
present, are the so called stylised empirical facts which constitute a
reference for realistic theoretical models. In connection with the fact
that the scaling range visible in the financial data typically extends
over only 1 - 1.5 order of magnitude, one has to keep in mind that the
stretched exponential distributions can also be considered reasonable
candidates~\cite{Laher} for modeling the financial fluctuations. Other
interesting related scenario is the one which corresponds to subordinated
stochastic processes~\cite{Clark} where time itself is a stochastic
process, or its multifractal~\cite{Mandelbrot1} and elastic
time~\cite{Dacorogna} generalizations.

From the point of view of the central limit theorem an essential element
is the speed of decay of correlations between the consecutive elementary
events. The speed of such a decay can be expected to be related to the
availability of information, opportunities to access it and quality of
its processing. These definitely systematically increase when going from
past to present which finds, for instance, evidence in a systematically
increasing frequency of trading. A natural question thus is to what
extent such elements can modify the dynamics of markets and, in
particular, if they can influence the characteristics mentioned above.

In addressing the related issues on the empirical level, we
systematically study the databases comprising the tick-by-tick price
changes of the 30 companies included in the Dow Jones Industrial Average
(DJIA)~\cite{TAQ} for most of the time during the period 1998-99, and of
the 30 companies included in the Deutsche Aktienindex (DAX)~\cite{Goeppl}
for most of the time during the same period. This corresponds to a
selection of stocks of similar market capitalization and thus their
dynamics compatible within either of these two groups, respectively.
Since we are dealing with a more recent history of the stock market
dynamics than the one presented in previous systematic analysis for the
American market by the Boston Group~\cite{Plerou} (years 1994-95), by
comparison, our present study can be oriented towards verifying the time
translational invariance of the relevant characteristics, of primary
interest being the probability of returns over varying time scales.
Secondly, such a selection of stocks also allows to compare the two
different stock markets in the same time intervals. Similarly as in
ref.~\cite{Plerou} the data from the TAQ databases have been filtered to
remove occasional spurious events.

\section{1998-1999 Stock Market Fluctuations}

When determining the distribution of returns, in order to obtain a
reasonable statistics, we consider fluctuations of all the companies
individually rather than those of the corresponding global index. The
resulting sample size (for the price changes sampled every 5 min) then
equals $30 \times 39000$ for the American market and $30 \times 52000$
for the German market. As it has been shown in ref.~\cite{Plerou}, the
fluctuations of the market and of its individual companies are typically
governed by distributions of essentially the same functional form and the
crossover to a Gaussian is even slower in the latter case (4 versus 16
days). For this reason the fluctuations of the companies are
expected~\cite{Plerou} to provide un upper bound for the distribution
characterising fluctuations of the global index.

\begin{figure}[t]
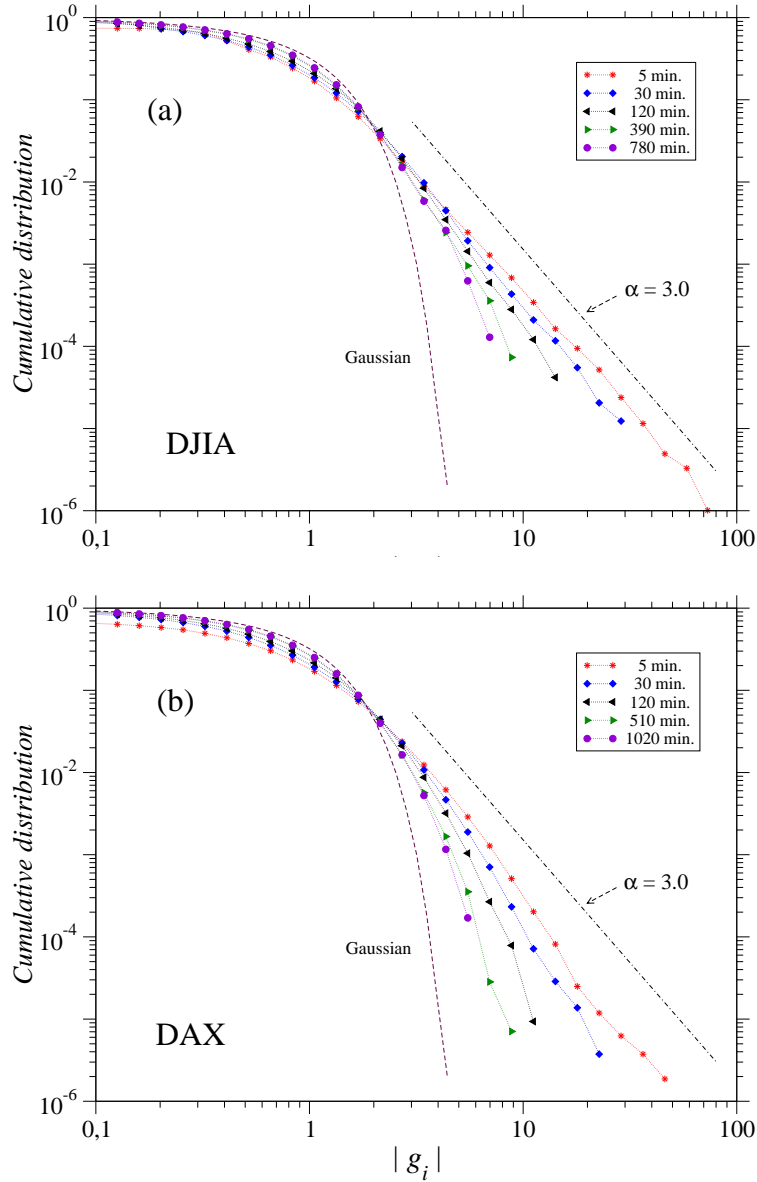

\hspace{1.2cm}
\epsfxsize 9.9cm
\epsffile{fig1a.eps}

%\vspace{-0.2cm}
\hspace{1.2cm}
\epsfxsize 9.9cm
\epsffile{fig1b.eps}
\caption{Cumulative distributions of the moduli of normalised returns of
the 30 companies which were included in the Dow Jones Industrial Average
(a) and of the 30 companies which were included in the Deutsche
Aktienindex (b) for most of the time during the same period 1998-99 
Different lines correspond to varying time scales $\Delta t$ starting 
from 5 min up to two trading days (780 min for DJIA and 1020 min for 
DAX).}
\label{fig:1}
\end{figure}

For the time series $P_i(t)$ representing the share price of $i$-th
company we use the commonly accepted definition of returns as
\begin{equation}
G_i \equiv G_i(t,\Delta t) = \ln P_i(t + \Delta t) - \ln P_i(t).
\label{G}
\end{equation}   
As another standard procedure, in order to make fluctuations of different
companies comparable, we make use of the normalised returns $g_i \equiv
g_i(t,\Delta t)$ defined as
\begin{equation}
g_i = {G_i - \langle G_i \rangle_T \over v_i},
\label{g}
\end{equation}
where $v_i \equiv v_i(\Delta t)$ of company $i$ is the standard deviation
of its returns over the period $T$
\begin{equation}
v_i^2 = \langle G_i^2 \rangle_T - \langle G_i \rangle_T^2
\label{v}
\end{equation}
and $\langle \dots \rangle_T$ denotes a time average. Since the
distribution of return fluctuations is typically to a good approximation
symmetric~\cite{Gopi,Plerou} with respect to zero, in the present
contribution we do not discuss such 'higher order' effects, and, in the
following, by returns we simply mean the moduli of returns.

\begin{figure}[t]
\hspace{1.2cm}
\epsfxsize 9.9cm
\epsffile{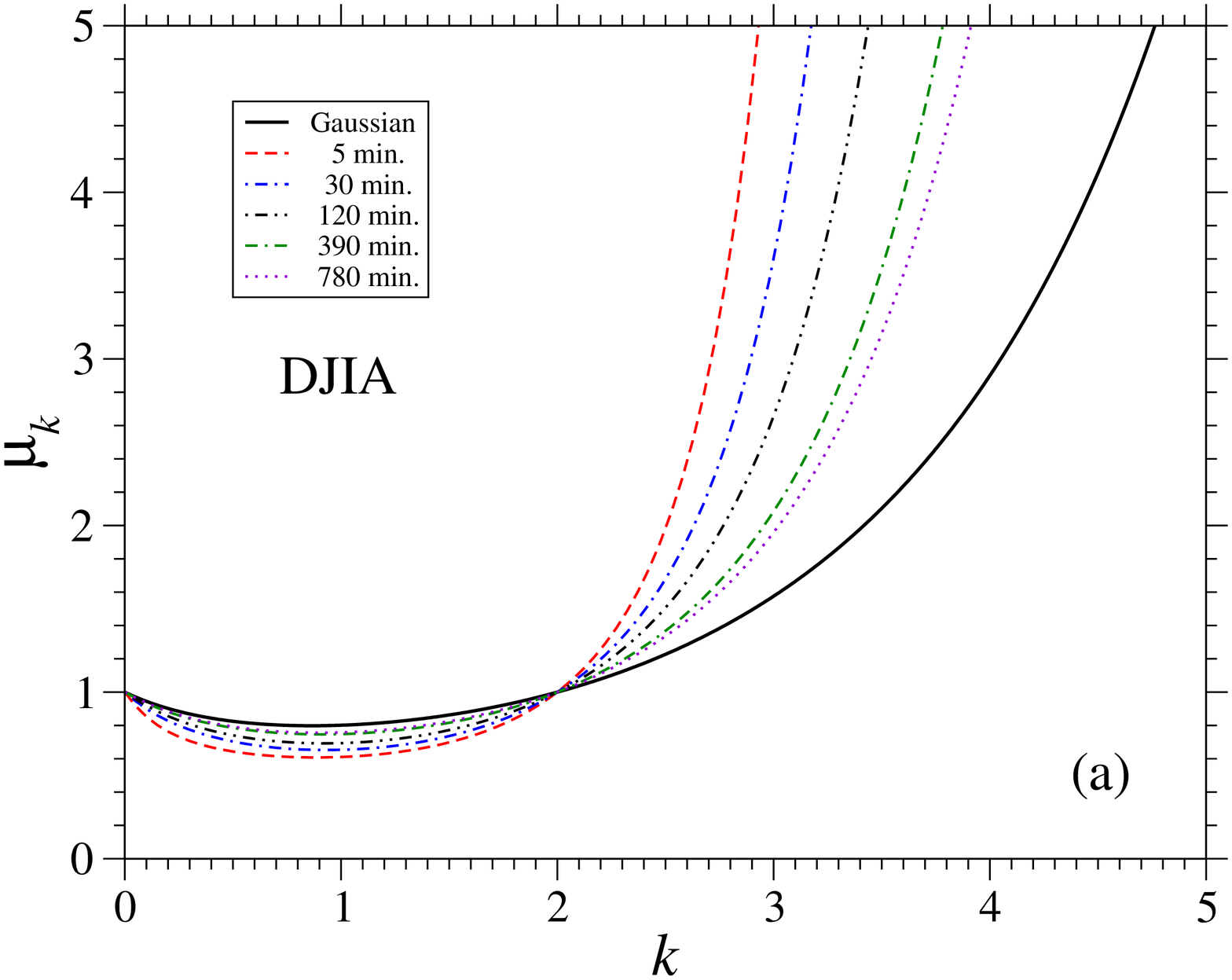}

%\vspace{0.4cm}
\hspace{1.2cm}
\epsfxsize 9.9cm
\epsffile{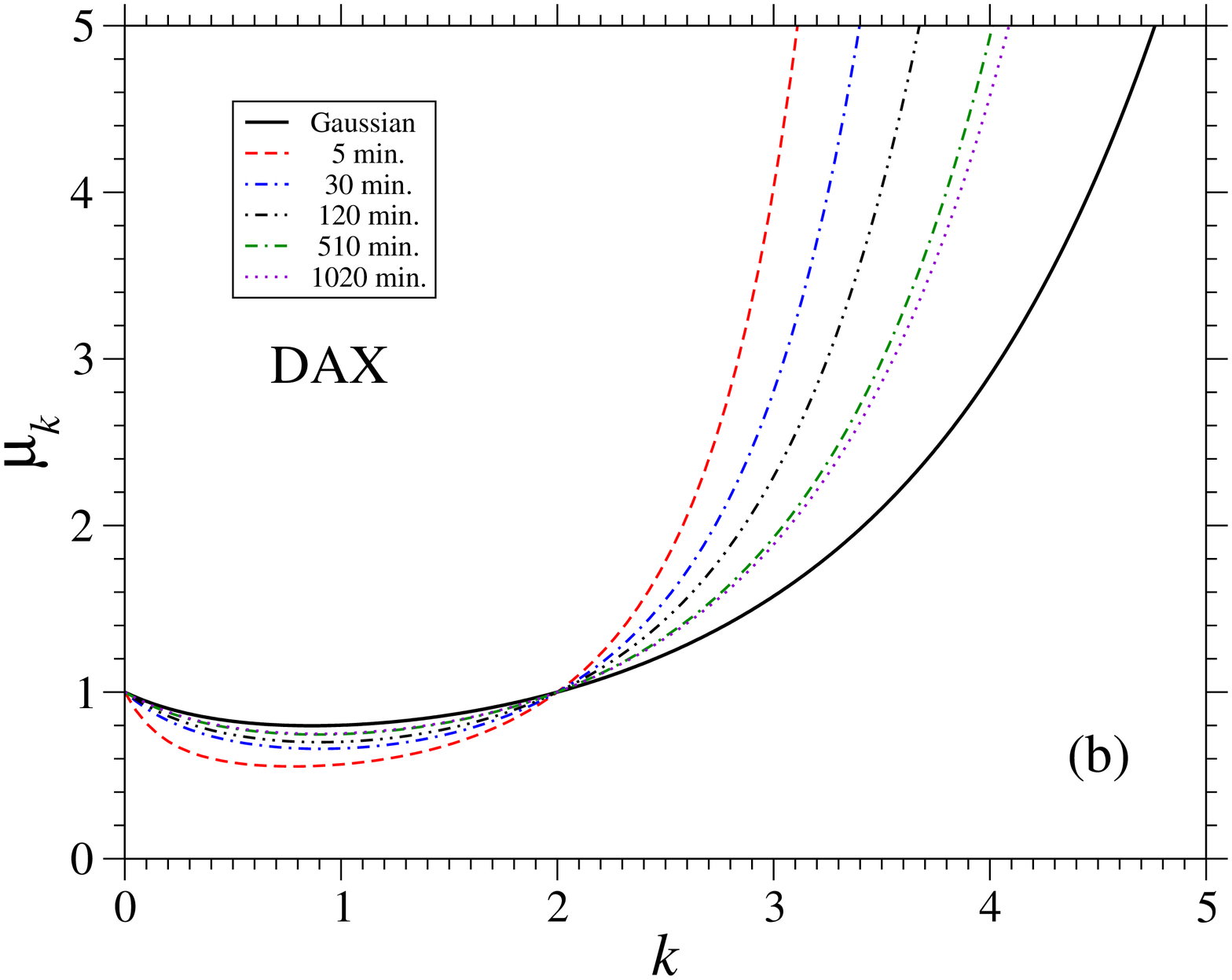}
\caption{Fractional moments for the normalised returns for the same cases
and for the same time scales as in Fig.~\ref{fig:1}. The solid full line
shows the Gaussian moments.}
\label{fig:2} 
\end{figure}

\begin{figure}[t]
\hspace{1.2cm}
\epsfxsize 9.9cm
\epsffile{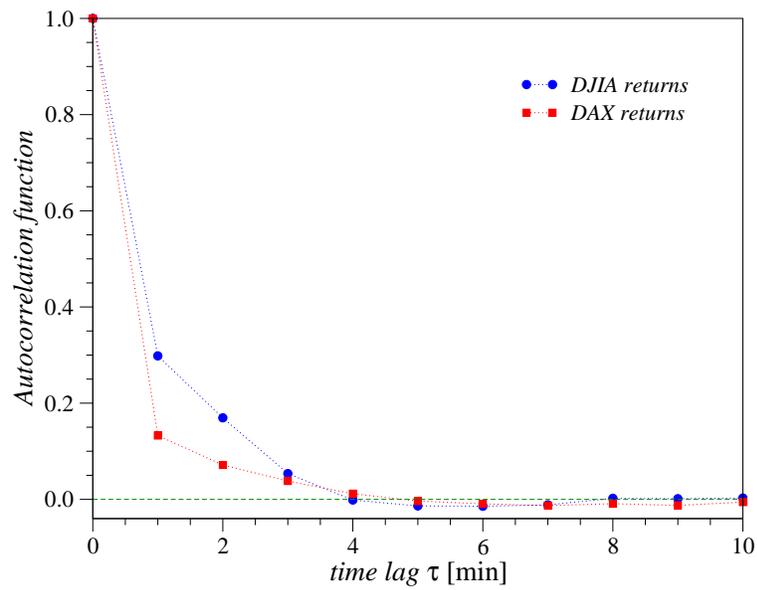}
\caption{Time-lag $\tau$ dependence of the autocorrelation functions
computed from the returns of the DJIA index and from the returns of
DAX index both sampled at a $\Delta t=1$ min time scale within the time
interval 1998-99.}  
\label{fig:3}
\end{figure}   

The cumulative distributions of such returns for the two sets of the
companies specified above are shown in Fig.~\ref{fig:1}. The most 
relevant here is their asymptotic behaviour which, based on the previous 
study, is expected to obey a power-law
\fussy
\begin{equation}
P(g > x) \sim x^{- \alpha},
\label{P}
\end{equation}
with $\alpha \approx 3$. The corresponding slope is indicated by the
dash-dotted line in this figure. On the short time scales ($\Delta t=5$
min and 30 min) the-DJIA-associated stock prices fluctuate according to
such a law, indeed. However, a deviation towards a Gaussian (dashed line)
can be seen starting already with $\Delta t=120$ min and it systematically
increases with increasing $\Delta t$. For the largest value of $\Delta
t=780$ min (two trading days for the DJIA) for which this characteristics
has been calculated, no scaling regime exists. The corresponding
transition in the case of the DAX companies turns out to occur even more
rapidly. In fact, in this case, already at $\Delta t=5$ min, the
distribution significantly deviates from $\alpha =3$ towards its larger
value. This is to be compared to a study~\cite{Lux} based on the older DAX
data which shows consistency with $\alpha=3$ for much larger time scales.
For the present data the fluctuations on the time scale of already one
trading day (for DAX this corresponds to 510 min) assume functional form
much closer to a Gaussian than to any scaling power-law.

\begin{figure}
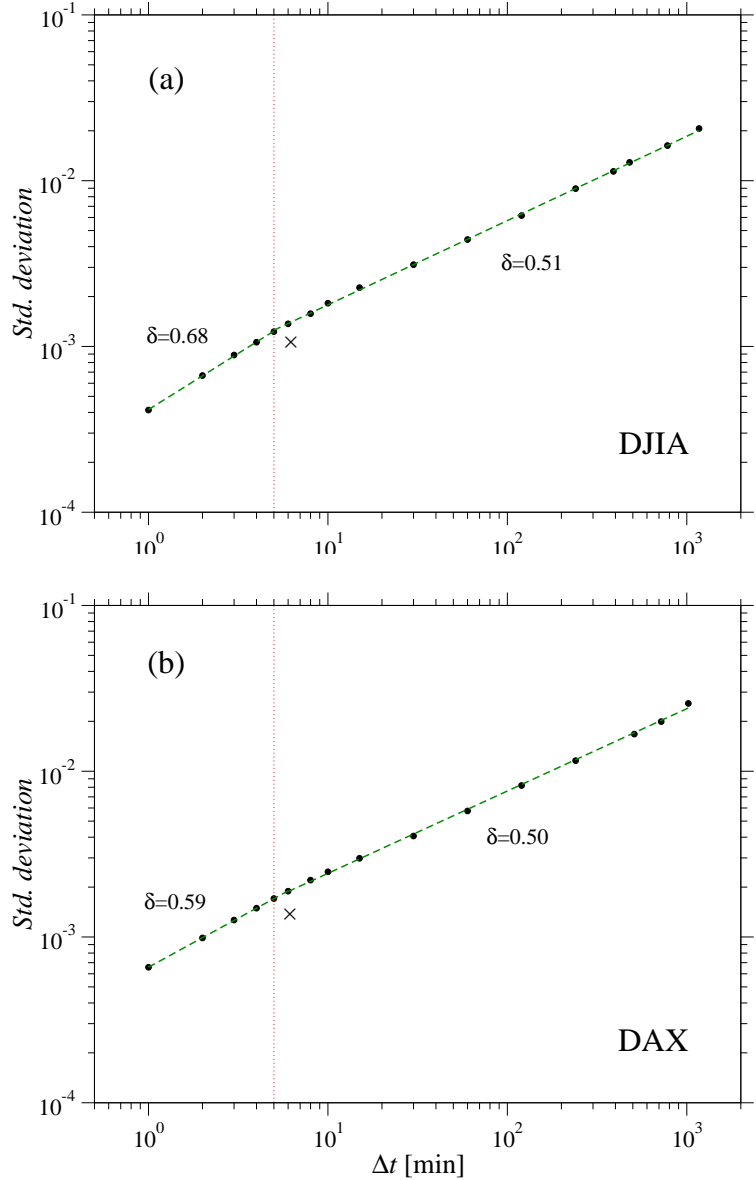
  
%\vspace{0.4cm}
\hspace{1.2cm}
\epsfxsize 9.9cm
\epsffile{fig4a.eps}

%\vspace{0.4cm}
\hspace{1.2cm}
\epsfxsize 9.9cm 
\epsffile{fig4b.eps}
\caption{(a) Time averaged volatility $v(\Delta t)$ as a function of the
time scale $\Delta t$ for the DJIA and (b) for the DAX within the same
time interval. Dashed lines represent fits in terms of $v(\Delta t)\simeq
\Delta t^{\delta}$. Vertical dotted lines indicate the crossover   
($\times$) at around $\Delta t=5$ min.}
\label{fig:4}
\end{figure} 

\begin{figure}[t]
\hspace{1.2cm}
\epsfxsize 9.9cm
\epsffile{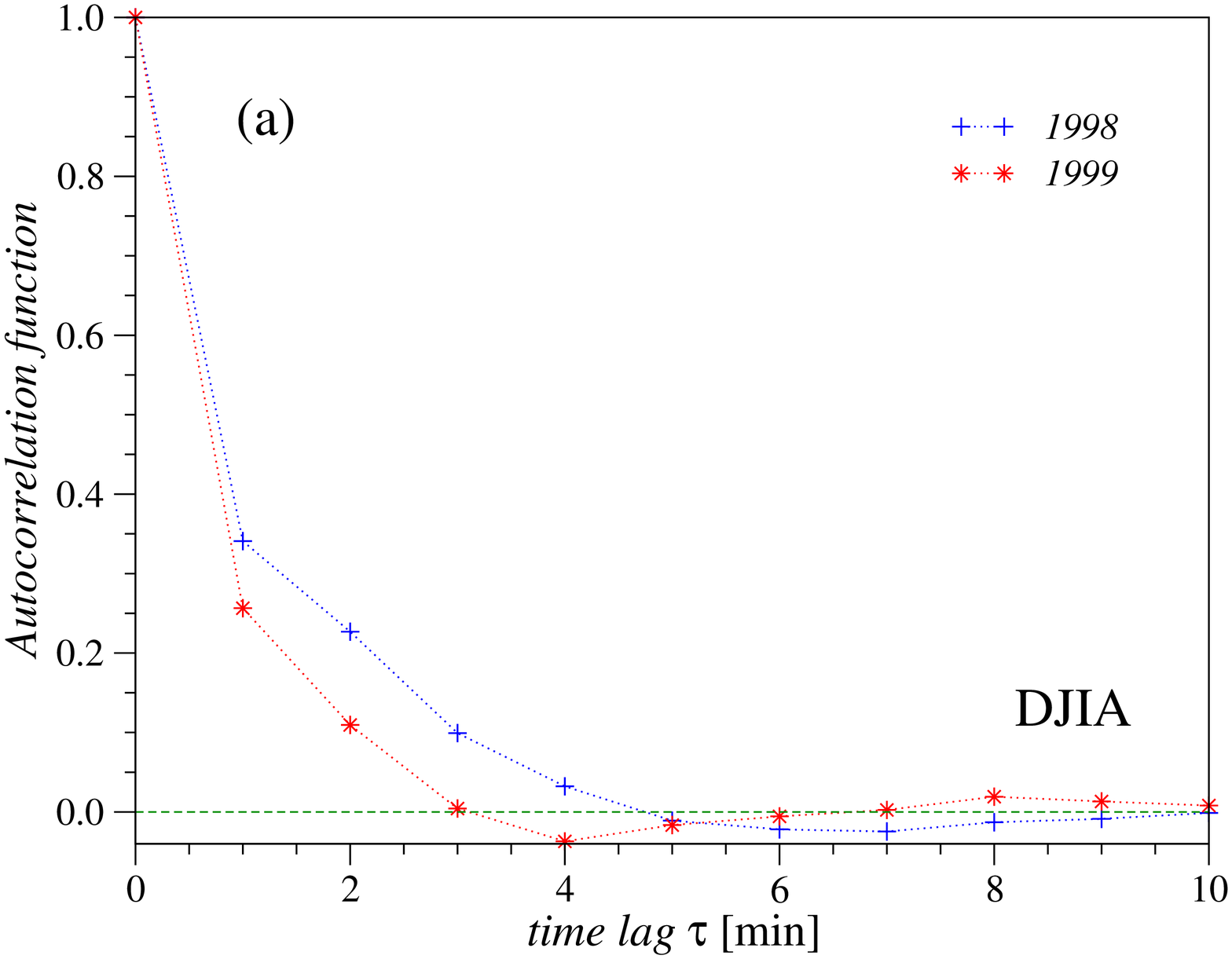}

\vspace{-0.2cm}
\hspace{1.2cm}  
\epsfxsize 9.9cm
\epsffile{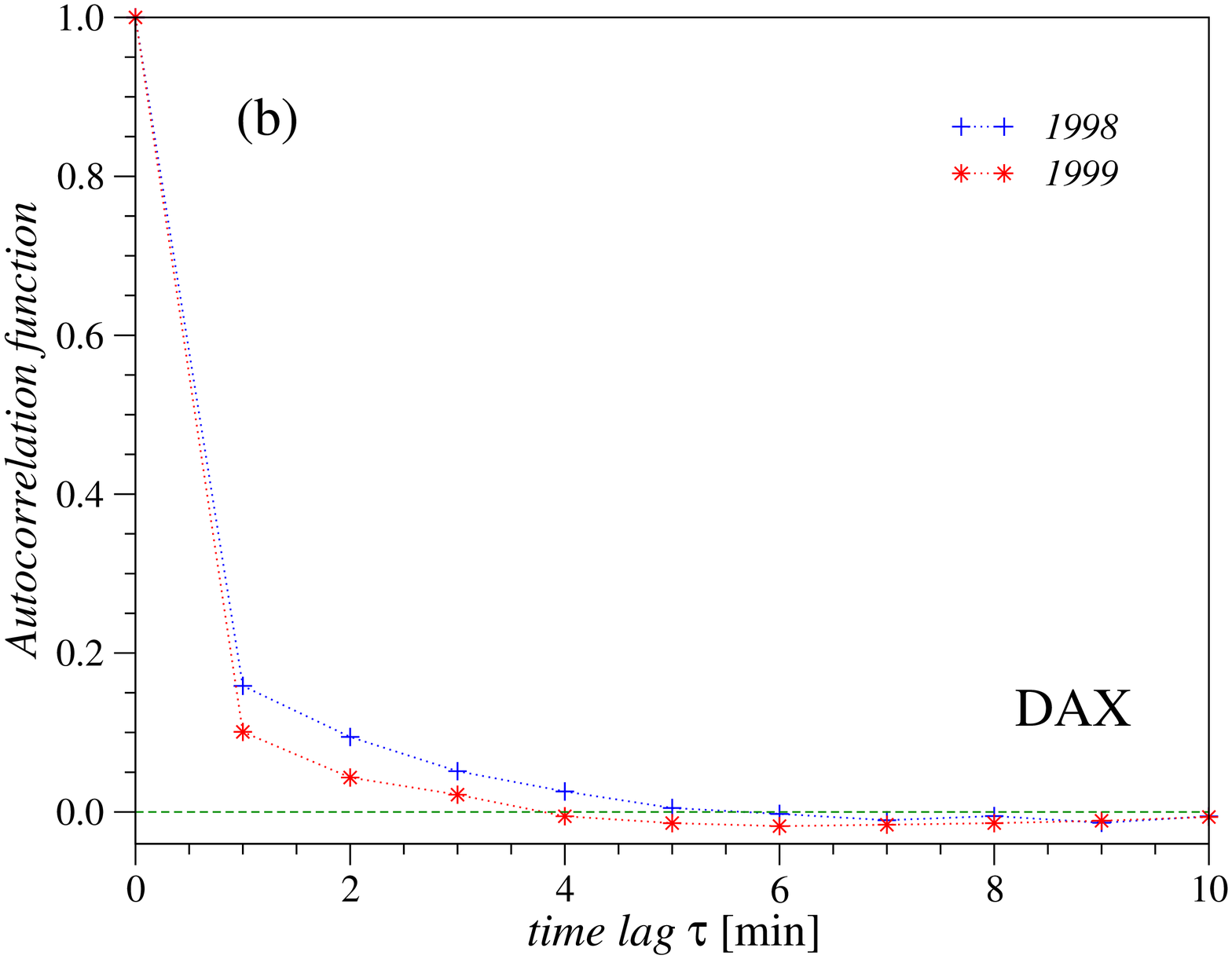}
\caption{Time-lag $\tau$ dependence of the autocorrelation functions of
returns for the DJIA (a) and for the DAX (b) returns sampled at a $\Delta
t=1$ min time scale within the time interval 1998 and 1999, separately.}
\label{fig:5}
\end{figure}

\begin{figure}[!ht]
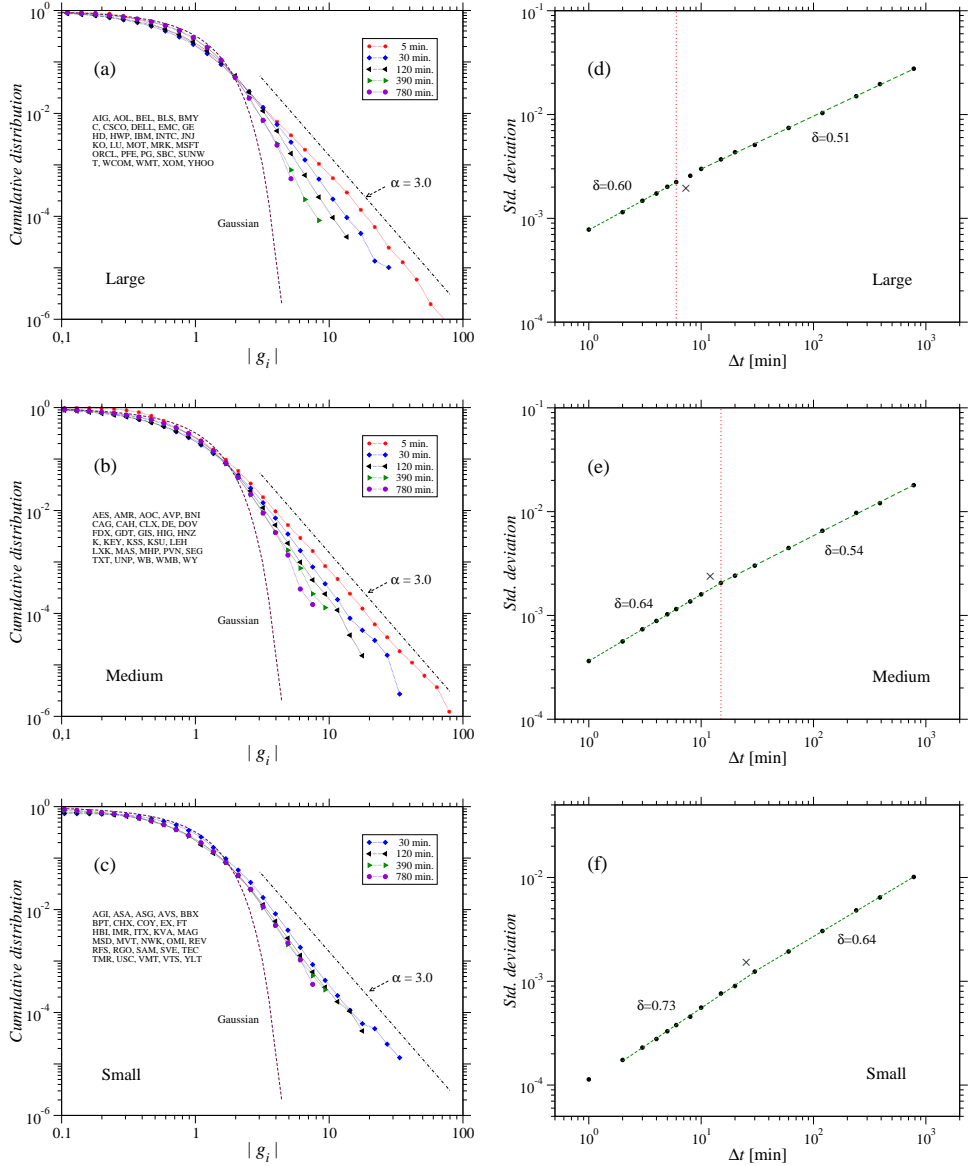

\hspace{-0.2cm}
\epsfxsize 6.2cm 
\epsffile{fig6a.eps}
\hspace{0.2cm}  
\epsfxsize 6.2cm
\epsffile{fig6d.eps}

\vspace{0.35cm}
\hspace{-0.2cm} 
\epsfxsize 6.2cm
\epsffile{fig6b.eps}
\hspace{0.2cm}
\epsfxsize 6.2cm
\epsffile{fig6e.eps}

\vspace{0.35cm}
\hspace{-0.2cm}
\epsfxsize 6.2cm
\epsffile{fig6c.eps}
\hspace{0.2cm}
\epsfxsize 6.2cm
\epsffile{fig6f.eps}

\caption{(LEFT) Cumulative distributions of the moduli of normalised
returns during the period 1998-99 of the three groups including 30
companies each, representing significantly different market
capitalisations $S$. These include (a) the largest ($S \ge 90$), (b)
medium ($10 \le S \le 15$) and (c) the lowest ($0.1 \le S \le 0.3$, all
in units of $10^9$ USD), available capitalisation, respectively.
Different lines correspond to varying time scales $\Delta t$ starting
from 5 min up to 780 min (two trading days). In (c) the time scale of 5
min is omitted due to a too large number of zero returns occurring in
this group of stocks. (RIGHT) Time averaged volatilities $v(\Delta t)$
for each of the three groups, correspondingly. In all these three cases
$v(\Delta t)$ is calculated from an ``index'' which is a sum of
split-adjusted prices of the 30 companies involved (d and e) and of 300
small companies (f).}
\label{fig:6}
\end{figure}

A more global quantitative measure of distributions is in terms of the moments.
For the normalised returns $g$ these are defined as
\begin{equation}
\mu_k = \langle \vert g \vert^k \rangle,
\label{m}
\end{equation}
and $\langle \dots \rangle$ denotes here an average over all the
normalised returns for all the bins. For both sets of returns the
so-calculated spectrum of moments is shown in Fig.~\ref{fig:2} for the 
same sequence of time scales as in Fig.~\ref{fig:1}. The moments can be 
seen to reflect basically the same tendency as it can be deduced from the
distributions of returns, i.e., a systematically increasing departure from
the $\alpha = 3$ scaling law in the region covered by the actual data.

A question now arises: is the above observation consistent with some more
dynamically oriented characteristics, like the autocorrelation function of
returns or the time averaged volatility $v(\Delta t)$ on different time
scales $\Delta t$? Indeed, an impressive consistency can be identified
when inspecting these characteristics shown in Figs.~\ref{fig:3} 
and~\ref{fig:4}, calculated here from the returns of the corresponding 
global indices, DJIA and DAX respectively, versus the behaviour of the 
distribution of returns from Fig.~\ref{fig:1}. The previous
study~\cite{Gopi,Plerou} shows that correlations in returns drop down to
the level of noise after about 20 min. In our case, this time is clearly
much shorter and equals about 5 min for both markets. This provides an
independent evidence that in the period 1998-99 the stock market
correlations cease to exist much faster than in the period 1994-95.
Interestingly, even though reaching the noise level after about the same 5
min, the speed of disappearence of correlations is larger for the DAX than
for the DJIA. This nicely correlates with the corresponding more abrupt
transition with increasing $\Delta t$ towards Gaussian (Fig.~\ref{fig:1}) 
in fluctuations of the DAX companies than those of the DJIA. It is also at
the same $\Delta t$ of 5 min where $v(\Delta t) \sim {\Delta t}^{\delta}$
changes its slope from superdiffusive $(\delta > 0.5)$ to normal $(\delta
= 0.5)$ for both markets. As consistent with behaviour of the
autocorrelation function, the dynamics of DJIA is more superdiffusive
$(\delta=0.68)$ in these initial 5 min than the one of the DAX.

In order to further illuminate on a possible origin of such a change of
the stock market dynamics we split our 1998-99 time interval into two
halves and for them separately calculate the autocorrelation functions of
returns. As shown in Fig.~\ref{fig:5}, we again can see an amazing
consistency for both markets: the more recent period of 1999 turns out to
be associated with a visibly faster decay of correlations than 1998, and
the autocorrelation functions for the whole period 1998-99, to a good
approximation, constitute the averages of the ones calculated over the
corresponding subintervals.

Finally, as an extra test of our analysis procedure and on the way
towards identifying further correlations between the above observations
and other measurable market characteristics, we select the three groups
from the TAQ database, including 30 companies each, representing
significantly different market capitalisations $S$. These include (a) $S
\ge 90$, (b) $10 \le S \le 15$ and (c) $0.1 \le S \le 0.3$, (all in units
of $10^9$ USD), i.e., the companies of the largest, medium and the lowest
capitalisation, respectively. The first group partially overlaps with the
DJIA. The corresponding cumulative distributions of returns for the same
different scales of time aggregation as before are shown in
Fig.~\ref{fig:6}(a)-(c). As it can be clearly seen the case (a) follows
the same tendency as the DJIA, the case (b) is somewhat less pronounced
in this respect but in the case (c) the slope of the distribution remains
essentially preserved up to the largest time scales considered.
Fig.~\ref{fig:6}(d)-(f) shows the time averaged volatilities $v(\Delta
t)$ for each of the above three groups, correspondingly. $v(\Delta t)$ is
here calculated from an ``index'' which is a sum of prices of the
companies involved. Summing up the prices is in fact close to the
price-weighted procedure of constructing the DJIA index. In the cases (d)
and (e) these are the same 30 companies listed in
Fig.~\ref{fig:6}(a)-(b), while in the case (f), in order to resolve the
dynamics down to the time scales of 1 min, the corresponding list of the
small companies is extended up to 300 (the small companies are
significantly less frequently traded which results in many zero 1 min
``index'' returns if a too small number of such companies is used). As
one can see, in the case of the largest companies $v(\Delta t)$ behaves
very similarly as for the DJIA itself (Fig.~4a), including the time scale
(5-6 min) of the transition from superdiffusive to normal. For the medium
size companies such a transition is somewhat delayed ($\sim$20 min) and
even not to a fully normal diffusion (from $\delta=0.64$ to
$\delta=0.54$). Continuing this way, for the small companies the dynamics
remains superdiffusive over the whole interval of the time scales
considered but still a transition from $\delta=0.73$ to $\delta=0.64$ can
be seen at around $\Delta t$=30 min. Again all this looks rather
consistent with the corresponding development of the distributions of
returns.

The analysis presented in Fig.~6 provides thus a test of significance of
the original (Figs.~1-4) results for the DJIA and for the DAX, since the
numbers of data points used are the same in all those cases. Secondly, in
view of the fact that an average frequency of transactions in the above
three groups of the companies is about (a) 15/min, (b) 1.5/min and (c)
0.2/min per company, correspondingly, it points just to this physical
parameter as the one which is directly related to the observed effects.
However, as a visible difference between the DAX and the DJIA in
approaching a limit of normal distributions shows, this definitely is not
the only relevant parameter. For the DAX the average number of
transactions per company is about 1/min and still it is DAX whose
departure from scaling and the decline of correlations in time is the
fastest among the cases considered here. A leading role of the DJIA in
dictating direction of the global stock market development has recently
been identified~\cite{Drozdz2} by studying correlation between the DAX
and the DJIA. Whether it is DAX which benefits from information already
preprocessed by the DJIA is an interesting possibility to be considered
in this connection.

\section{Conclusions}

These results provide quite a remarkable indication that the contemporary
financial dynamics on average is more efficient in the sense of the
efficient market hypothesis~\cite{Fama} in its weak form, as compared to
a more distant history. From the practical point of view this may be
considered good news for the conventional option pricing
methods~\cite{Black,Merton} which assume a normal distribution of
financial fluctuations. In a sense this result also provides some more
arguments in favour of the standard extreme value theory~\cite{Embrechts}
for estimating the value-at-risk for very low probability extreme events.
The related literature assumes independent returns which implies the
decreasing degree of fatness in the tails. There is still one more
element that is to be kept in mind when trying to interpret the present
observations. The world stock markets, including the two considered here,
were experiencing more sizable increases during the period 1998-99 than
during 1994-95. As shown in ref.~\cite{Drozdz1}, such periods are
typically more noisy and more competitive as far as correlations among
the individual stocks are concerned. Just a time-translation is thus not
the only element when relating those two periods of the stock market
history. In any case, however, the issue of the so-called financial
stylised facts needs to be revised and, possibly, generalised to
incorporate an increasing access to information and ability to process it
when going from past to present. All this provides further arguments for
being time-adaptive, and even market-adaptive, when looking into the
dynamics of the financial markets, which is especially important for an
appropriate perception of the risk involved.

S.D. acknowledges support from Deutsche Forschungsgemeinschaft (DFG)
under contract Bo 56/160-1. This work was also supported in part by DFG
grant no. 447 Aus 113/14/0.

J.K. and J.S. thank Tony Thomas for valuable discussions and the 
hospitality they enjoyed in the CSSM where a part of this article was
written.

\end{document}